\documentstyle[aps,prc,epsfig,preprint]{revtex}
\tightenlines

\begin{document}
\draft
\begin{titlepage}
\title{Thermodynamical properties of a mean-field plus pairing 
       model and applications for the Fe nuclei 
       }
\author{S. Rombouts \thanks{Postdoctoral Fellow of the Fund for Scientific 
                            Research - Flanders (Belgium)},
        K. Heyde \thanks{present address: EP-Isolde CERN, CH 1211, Geneva 23}
        and N. Jachowicz}
\address{Universiteit Gent,
         Vakgroep Subatomaire en Stralingsfysica
         \\
         Proeftuinstraat 86, B-9000 Gent, Belgium
         \\
         E-mail: Stefan.Rombouts@rug.ac.be, Kris.Heyde@rug.ac.be
         }
\date{April 27, 1998}
\maketitle
\begin{abstract}
\noindent
A mean-field plus pairing model for atomic nuclei in the Fe region 
was studied using a finite-temperature quantum Monte-Carlo method. 
We present results for thermodynamical quantities 
such as the internal energy and the specific heat.
These results give indications of a phase transition related to the pairing 
amongst nucleons, around temperatures of 0.7 MeV.
The influence of the residual interaction and of the size of the model space 
on the nuclear level densities is discussed too.
\end{abstract}
\vspace{1cm}
\centerline{To appear in Physical Review C.}
\vspace{1cm}

\end{titlepage}
%
%
\section{introduction}
Quantum Monte-Carlo methods offer an interesting way to study 
fermionic many-body problems. 
Accurate calculations of ground-state 
properties have been performed for light nuclei,
using Variational and Diffusion Quantum Monte-Carlo methods \cite{pudliner}. 
Also for the nuclear shell model, Monte-Carlo methods are very useful
\cite{smmc}.
They allow to do calculations in much larger model spaces than 
conventional techniques based on diagonalization.
Furthermore, they are very useful for the study of finite-temperature 
properties of atomic nuclei \cite{deanfe,pairsmmc1}.
The same method can also be used as a starting point to
calculate nuclear level densities \cite{ormandl,nakada}.
We have used a variant of this method to study 
thermodynamical properties of nuclei in the Fe region 
with a model based on a mean-field plus pairing Hamiltonian.
At present, we are limited to this too simple model, because our 
quantum Monte-Carlo method is still in a developing phase and 
because we are limited by our present computer facilities 
(a Dec Alphastation 255/300MHz Workstation and a 200 MHz Pentium PC).
Even though, at present, our results are not conform to realistic 
Hamiltonians, they do say a lot about the general physics properties.
It is a first step to go beyond the mean field approximation. 
The present approach gives results about the possibility of a phase transition
related to pairing correlations,  
about the influence of the residual interaction on the level densities
and on a number of general features of the nuclear many-body structure 
at finite temperature.

The combination of a mean-field potential and the pairing  Hamiltonian 
leads to a Hamiltonian $\hat{H}=\hat{H}_{\rm mf}+\hat{H}_P.$
Though this Hamiltonian looks simple, 
it already leads to a complicated many-body problem.
An often used technique to tackle this problem,
is the Bardeen-Cooper-Shrieffer (BCS) theory \cite{ring}.
It leads to equations that can be handled easily in a numerical way
and it has a clear interpretation in terms of quasiparticles.
The disadvantage is that it gives only an approximate solution,
and leads to many-body states in which the number of particles is not fixed.
For some systems, exact solutions can be found \cite{richardson}.
A general, accurate  solution for this many-body problem is 
even at present a topic of intensive research \cite{burglin,cerf2}.
We have found that our Quantum Monte-Carlo method is a very useful method 
in order to study the ground-state and finite-temperature 
properties of the nuclear pairing Hamiltonian.

In Section \ref{sectionmodel} we introduce the mean-field plus pairing model 
that we used for our calculations. 
In Section \ref{sectionqmc} aspects of our Quantum Monte-Carlo method are 
given, with an emphasis on the differences with related methods. 
In Section \ref{sectionresults} we present a number of results. 
The role of the pairing-interaction strength, the size of the model space 
and the number of particles is discussed. 
An estimate is given for the level densities. 
The lines that connect the data points on the figures in this paper
are ment to guide the eye. 
They do not correspond to analytical results nor fitted curves,
except for the pure mean-field results (G=0).
Error limits represent $95\%$ confidence intervals.
If no error limits are shown, 
it means that they are smaller than the markers of the data points,
unless it is stated that no error limits were determined.
\newpage
\section{A mean-field plus pairing model for the Fe nuclei.}
\label{sectionmodel}

For the mean-field potential, a Woods-Saxon potential $U(r)$ is used 
\cite{woods}:
\begin{eqnarray}
\label{woodssax}
 U(r) &=& V_c - V f(x) 
          + \left(\frac{\hbar}{m_{\pi} c} \right)^2 V_{so} \left({\bf \sigma}
 \cdot {\bf l} \right)
               \frac{1}{r}\frac{d}{dr} f(x_{so}),
 \\ {\rm where } & & \nonumber \\
  V_c &=& Z e^2/r, \ \ \ r \geq R_c, \nonumber \\
      &=& \left[Z e^2/(2R_c) \right]  (3 - r^2/R_c^2) , \ \ \ r \leq R_c,   
 \\
  R_c &=& r_c A^{1/3}, \nonumber \\
  f(x)&=& (1+e^x)^{-1} \ \ {\rm with} \ x= \left( r -r_0 A^{1/3} \right) /a, 
 \\
   \left(\frac{\hbar}{m_{\pi} c} \right)^2 &=& 2.000 \ fm^2.
\end{eqnarray}
Here, $A$ is the number of nucleons, $Z$ the number of protons.
The other parameters are taken as in \cite{perey}:
\begin{eqnarray*}
  r_0    &=& 1.25 \ fm, \ \ \
  a      = 0.65 \ fm, \ \ \
  V      = 53.3  + 27(A-2Z)/A -0.4 Z/ A^{1/3} \ MeV, \\
  r_{so} &=& 1.25 \ fm,  \ \ \
  a_{so} = 0.47 fm, \ \ \
  V_{so} = 7.5 \ MeV.
\end{eqnarray*}
To calculate the mean field and its eigenfunctions,
we use the parameters for the nucleus $\mbox{}^{56}_{26}\mbox{Fe}_{30}$.
This mean field is used for all nuclei in this particular mass region.
For every set of quantum numbers $l$ and $j$, 
the Woods-Saxon potential is diagonalized in a basis of 
the lowest 60 harmonic-oscillator eigenfunctions with the appropriate symmetry.
In this way, the single-particle eigenstates and their energies 
listed in table \ref{table1}, are obtained.
Also a number of unbound states (with energy $E>0$) are obtained.
In fact, the Woods-Saxon potential exhibits a continuum of unbound eigenstates.
Due to the expansion in a finite number of basis functions,
discrete unbound energy levels are obtained.
These can be seen as a discrete approximation to the continuum of unbound 
states. 
The $1s_{\frac{1}{2}}$, $1p_{\frac{3}{2}}$, $1p_{\frac{1}{2}}$,
$1d_{\frac{5}{2}}$, $1d_{\frac{3}{2}}$ and $2s_{\frac{1}{2}}$ orbitals 
are considered to be compeletely filled.
They form an inert core for the many-body problem.
The $ 1f_{\frac{ 7}{2}} $, $ 2p_{\frac{ 3}{2}} $,$ 2p_{\frac{ 1}{2}} $ 
and $ 1f_{\frac{ 5}{2}} $ orbitals constitute the valence shell.

A simple Hamiltonian that accounts for the short-range correlations 
induced by the residual interaction,
is the nuclear pairing Hamiltonian $\hat{H}_P$ \cite{ring}, 
that takes the form
\begin{equation}
\label{pairingham}
 \hat{H}_P=  - \sum_{t=p,n} G_t \sum_{k,k'>0}
                   \hat{a}_{k' t}^{\dagger} \hat{a}_{\bar{k't}}^{\dagger}
                   \hat{a}_{\bar{k t}}\hat{a}_{k t},
\end{equation}
where the operators $\hat{a}_{k t}^{\dagger}$ create a particle in 
the corresponding single-particle eigenstates 
of the mean-field Hamiltonian in the valence shell.
The index $t$ indicates proton or neutron states,
$\bar{k}$ is the time-reversed state of the state $k$.
The notation $k,k'>0$ denotes that the summation for $k$ and $k'$ 
should run over states with angular momentum projection  $j_z>0$ only.
The interaction strength $G_t$ depends on the model space and 
the system under study.
For the strength of the pairing interaction we take $G=20MeV/56$, following 
the suggestion of Bes and Sorensen \cite{bes} to take $G=20Mev/A$.
The same strength is used for protons and neutrons,
and for all nuclei in the Fe mass region.
\newpage

\section{The quantum Monte-Carlo method}
\label{sectionqmc}
The method we used to study the model is based on the Shell-Model 
Quantum Monte-Carlo method. 
The basic idea of this method is well explained in reference \cite{smmc}.
It amounts to a decomposition of the Boltzmann-operator $e^{-\beta \hat{H}}$
in a sum of exponentials of one-body operators 
\begin{equation}
\label{edecomp}
e^{-\beta \hat{H}}=\sum_{\sigma} e^{- \hat{h}_{\beta,\sigma}}.
\end{equation}
Because of their one-body nature, the terms $e^{-\hat{h}_{\beta,\sigma}}$
can be handled easily using small matrices, with a dimension
equal to the number of single-particle states considered in the model space.
The canonical or grand canonical trace of $e^{-\hat{h}_{\beta,\sigma}}$ 
can be calculated with simple algebraic operations on these matrices.
The sum over the {\it auxiliary fields} $\sigma$ is then evaluated using
Monte-Carlo techniques (in casu the Metropolis algorithm \cite{metropolis}).

In our approach, a number of technical aspects are implemented
differently from reference \cite{smmc}. The main differences are:
\begin{itemize}
\item 
In order to arrive at a decomposition of the form \ref{edecomp}, we use the 
Suzuki-Trotter formula \cite{suzuki} to separate the one- and two-body part 
of the Hamiltonian in the exponents. 
This reduces the leading systematic error term to order $\beta^3/{N_t^2}$, 
with $N_t$ the number of inverse-temperature slices. 
Following the prescriptions from reference \cite{smmc}, 
the leading systematic error term would be of the order $\beta^2/{N_t}$.
\item 
In order to decompose the exponential of the two-body part of the Hamiltonian,
we use the discrete Hubbard-Stratonovich, described in \cite{dischubs}.
Compared to the Hubbard-Stratonovich transform described in 
\cite{smmc,hubbard,straton},
it has the advantage that it leads to faster matrix operations and to  smaller
systematic errors. Because of these fast matrix operations, we could use a 
large number of inverse-temperature slices in order to reduce the systematic 
errors (typically $N_t=20\beta$ for the $f p$-model space and $N_t=40\beta$ for
the extended model spaces).
\item 
In order to evaluate the canonical trace of $e^{-\hat{h}_{\beta,\sigma}}$, 
we use the fast an efficient algorithm described in reference \cite{tracec} 
instead of the number projection technique described in \cite{smmc,ormand}.
\item 
The trial steps for the Metropolis algorithm for the sampling of the 
discrete auxiliary fields $\sigma$ are generated in the following way:
a series of $M$ consecutive inverse-temperature intervals are 
chosen randomly to be updated. 
Because of the permutation properties of the canonical trace,
we can always shift these inverse-temperature intervals to the end of
the decomposition. The matrix representation $U_\sigma$ of
the operator $e^{-\beta \hat{h}_{\sigma}}$ is calculated up to the 
$(N_t-M)^{th}$ inverse-temperature slice and stored in computer memory.
A number $m$ between 1 and a maximum number is drawn.
Then in the last $M$ slices $m$ auxiliary fields are drawn. 
These are then changed randomly. 
The matrix $U_{\sigma'}$ for the altered configuration $\sigma'$ is
constructed out of the stored part of $U_{\sigma}$.
The canonical trace of $U_{\sigma'}$ is evaluated. 
Then the configuration $\sigma$ is accepted or rejected according to
the Metropolis algorithm \cite{metropolis}.
This procedure is repeated a number of times (typically 7 times),
before a new series of $M$ slices is selected.
So a complete Markov step consists of 7 local updates. 
Typical values are $M=80$ (or $M=N_t$ for $N_t<80$)
and $1< m < 160$.
This scheme allows to update a large number of auxiliary fields simultaneously,
while requiring only $M$ matrix multiplications per udpate
(counting the contribution of one inverse-temperature slice as one matrix).
Only after a complete Markov step all the $N_t$ 
matrices in the decomposition have to be multiplied.

\item 
Observables are evaluated after every 5 complete Markov steps.
The values are not yet fully decorrelated at this rate
(e.g. autocorrelations between 30\% and 60\% between consecutive
values for the energy), 
but leaving a larger interval would not improve the performance, 
because already at this rate the most time-consuming
part is the construction of the trial configurations for the Markov chain.
The value of an observable $\hat{A}$ at inverse temperature $\beta=1/(kT)$
is calculated as follows:
\begin{eqnarray}
\langle \hat{A} \rangle_{\beta} &=&
\frac{\mbox{Tr}_{N} \left(\hat{A}  e^{-\beta \hat{H}} \right) }
     {\mbox{Tr}_{N} \left(e^{-\beta \hat{H}} \right) },
\label{exprobs1} \\ & = &
\frac{\left. \frac{d}{d \epsilon} 
\mbox{Tr}_{N} \left(e^{-\beta \hat{H}+\epsilon \hat{A}} \right) 
\right|_{\epsilon=0} }
     {\mbox{Tr}_{N} \left(e^{-\beta \hat{H}} \right) },
\label{exprobs2} \\ & = &
\mbox{E}_{w} (A),
\label{observs}
\\ \nonumber {\rm with} & & \\
\mbox{E}_{w} (A) &=&
\frac{ \sum_{\sigma} A_{\sigma} w_{\sigma}} { \sum_{\sigma'} w_{\sigma'}},
\\ 
A_{\sigma} &=& \left. \frac{d}{d \epsilon} \log \left[
\mbox{Tr}_{N} \left(e^{\hat{h}_{\beta,\sigma,\epsilon}} \right) \right] 
\right|_{\epsilon=0}, \\
w_{\sigma} &=& \mbox{Tr}_{N} \left(e^{\hat{h}_{\beta,\sigma}} \right) .
\end{eqnarray}
Here, $\mbox{Tr}_N$ denotes the canonical trace, i.e. the trace over all 
$N$-particle states.
The mathematical properties of the trace operator are crucial in going 
from expression \ref{exprobs1} to expression \ref{exprobs2}.
The one-body operator
resulting from the decomposition \ref{edecomp}, with $- \beta \hat{H}$
replaced by $-\beta \hat{H} +\epsilon \hat{A}$,
is denoted by $\hat{h}_{\beta,\sigma,\epsilon}$.
For the energy ($\hat{A}=\hat{H}$), this can easily be implemented as
$ \hat{h}_{\beta,\sigma,\epsilon}=\hat{h}_{\beta-\epsilon,\sigma}$.
The derivatives are evaluated as
\begin{equation}
  \left. \frac{d}{d \epsilon} f(\epsilon) \right|_{\epsilon=0} \simeq
  \frac{f(\epsilon_0)-f(-\epsilon_0)} {2 \epsilon_0},
\end{equation}
with $\epsilon_0$ a small but finite number (typically $\epsilon_0=1/2048$).
This way of evaluating observables requires a lot of matrix manipulations
because the complete matrix $U_{\sigma}$ has to be recalculated from scratch
for $\hat{h}_{\beta, \sigma, \epsilon_0}$ 
and $\hat{h}_{\beta, \sigma,-\epsilon_0}$.
However, because most of the computing time goes to 
the construction of the trial moves, 
this has little impact on the overall performance.
Furthermore, this way of evaluating the observables leads to small
statistical errors, because it amounts to an insertion of the operator 
$\hat{A}$ at each inverse-temperature interval,
whereas the procedure described in \cite{smmc} is based on the 
insertion of $\hat{A}$ only at the first interval.
A special remark concerns the evaluation of the specific heat $C$.
This quantity cannot be calculated as the expectation value of an observable.
We evaluate $C$ after the Monte-Carlo run as 
\begin{eqnarray}
\label{exprobsc}
C & =& \beta^2 \left(\mbox{E}_{w} (E_2) - \mbox{E}_{w} (E)^2 \right), \\
\nonumber with  & & \\
E_\sigma &=& - \frac{d}{d \beta} 
\log \left[\mbox{Tr}_{N} \left(e^{\hat{h}_{\beta,\sigma}} \right) \right],
\nonumber \\
E_{2 \sigma} &=&
 \frac{d^2}{(d \beta)^2} 
\log \left[\mbox{Tr}_{N} \left(e^{\hat{h}_{\beta,\sigma}} \right) \right]
+ E_{\sigma}^2.
\nonumber
\end{eqnarray}
The observable $E_2$ corresponds to the square of the Hamiltonian:
\begin{equation}
  \mbox{E}_{w} (E_2) 
  =\mbox{Tr}_N \left(\hat{H}^2 e^{-\beta \hat{H}} \right) / 
   \mbox{Tr}_N \left( e^{-\beta \hat{H}} \right) . 
\end{equation}
\item 
Just like any other Quantum Monte-Carlo method for fermions \cite{fahy,lind}, 
this method suffers from a sign  problem at low temperatures.
The value of the weights $w_{\sigma}$ can become negative.
This poses a problem for the Metropolis algorithm, because it requires
that $w_{\sigma}$ can be interpreted as a probability density.
The Metropolis algorithm can still be used by applying it to the absolute
value $|w_{\sigma}|$. 
Then we have to treat the sign $s_{\sigma}$ of $w_{\sigma}$ as an observable.
Expression \ref{observs} becomes:
\begin{eqnarray}
\langle \hat{A} \rangle_{\beta} & = &
\frac{ \sum_{\sigma} A_{\sigma} s_{\sigma} |w_{\sigma}|} 
     { \sum_{\sigma'} s_{\sigma} |w_{\sigma'}|},
 \nonumber \\
&=& \frac{\mbox{E}_{|w|} (A s)} {\mbox{E}_{|w|} (s)}.  
\end{eqnarray}
The problem is that the statistical error on this expression scales 
as $1/  \mbox{E}_{|w|} (s)$ (see below).
Now, for an even number of protons and an even number of neutrons,
one can exploit a symmetry between states with 
$j_z > 0$ and states with $j_z < 0$.
This symmetry guarantees that $\mbox{E}_{|w|} (s)$ is close
to 1 even at low temperature \cite{dischubs}.
In reference \cite{smmc} it was claimed that the canonical trace should be 
exactly equal to one for even particle numbers with interactions that lead to 
matrix representations of the form 
\begin{equation}
\label{signmatrix}
U_{\sigma} = \left( \begin{array}{cc}  P & Q \\ -Q^{*} & P^{*} 
                         \end{array} \right),
\end{equation}
as is the case with a pure pairing force or a pairing-plus-quadrupole force. 
This is not exactly true however, as can be seen from the following 
counterexample: let $U_{\sigma}$ be a $4 \times 4$ matrix, 
$P$ and $Q$ are $2 \times 2$ matrices, with $Q=0$ and
\begin{equation}
P=  \left( \begin{array}{cc}  \epsilon  & 0 \\ 0 & -\epsilon
                         \end{array} \right), 
\end{equation}
where $\epsilon$ is a real number. 
Then the grand canonical trace is given by
\begin{eqnarray}
\nonumber
\mbox{Tr}_{\mu} \left( \hat{U}_{\sigma} \right) & = &
 \left(1- \epsilon^2 e^{2 \beta \mu} \right)^2 \\
& = & 1- 2  \epsilon^2 e^{2 \beta \mu} + \epsilon^4 e^{4 \beta \mu}, 
\end{eqnarray}
which is positively definite for any value of the chemical potential $\mu$.
The canonical 2-particle trace is given by the coefficient 
of $e^{2 \beta \mu}$: 
\begin{equation}
\mbox{Tr}_{2} \left( \hat{U}_{\sigma} \right) = - 2 \epsilon^2 ,
\end{equation}
which is negative.
The average sign for some of the systems we studied is shown in figure 
\ref{figurepairnsigns} as a function of the inverse temperature $\beta$.
The sign for the even particle numbers shows a dip around $\beta=2 MeV^{-1}$.
This is caused by the fact that for a fraction of the $U_{\sigma}$ 
the canonical trace becomes negative, like in the above example. 
However, the symmetry between $j_z > 0$ states and  $j_z < 0$ states
guarantees that the contribution to the trace of the 'fully accompagnied' 
states, in the sense defined in reference \cite{burglin}, will always
be positive. These states dominate the low temperature trace for even-even 
nuclei. Therefore the sign  tends to one again at low temperatures.
For odd particle numbers, $\mbox{E}_{|w|} (s)$ tends to zero
at low temperatures. 
So there the statistical errors explode.
We find that even for odd nuclei we can do calculations
at $\beta = 4 MeV^{-1}$, which is enough to cool the system
almost completely to its ground state. 
For the model studied here, the average sign posed no problems. 
However, for more complicated Hamiltonians, one probably will have 
to fall back on extrapolation techniques as the ones described in 
references \cite{smmc} and \cite{alhassid}, 
in order to overcome the sign problem.
\item 
Because the Metropolis algorithm leads to correlations amongst successive
values for the observables, care has to be taken to establish
accurate error limits.
In order to get rid of the correlations, 50 independent Markov chains are
run for each calculation.  
These Markov chains typically consist of some 600 thermalization steps and 
3000 sampling steps, with an evaluation of observables every $5^{th}$ step.
 
So the chains are rather short, but long enough to make sure that 
the computing time is not dominated by the thermalization steps.
This leads to 50 independent estimates $A_1, \ldots, A_{50}$ for 
the quantity $\langle \hat{A} \rangle_{\beta}$.
To obtain the final estimate, we take the weighted average of these
values, with the average signs  $s_1, \ldots , s_{50}$ as weights.
So we obtain an estimate $\bar{A}$ for $\langle \hat{A} \rangle_{\beta}$:
\begin{equation}
\label{weighteda}
 \bar{A} = \frac{\sum_i A_i s_i} {\sum_j s_j}.
\end{equation}
If the $A_i$ and $s_i$ are obtained with enough precision,
then a good estimate for the statistical error on \ref{weighteda}
can be obtained from the following expression: \cite{hastings}
\begin{equation}
 \mbox{var}(x/y) =\left\{ \mbox{var} (x) - 2 \mbox{E} (x/y) \mbox{cov} (x,y)
      + \mbox{E}^2 (x/y) \mbox{var} (y) \right\} / \mbox{E}^2 (y),
\end{equation}
($\mbox{var}$ and $\mbox{cov}$ denote the variance and covariance).
Taking $x=As$ and $y=s$, 
leads to an estimate for the variance on $\bar{A}$:
\begin{equation}
\label{avar}
 \mbox{var} (\bar{A}) \simeq
 \frac{ \sum_{i} (A_i - \bar{A})^2 s_i^2 } 
      {\left( \sum_j s_j \right)^2}. 
\end{equation}
Under the assumption that $\bar{A}$ is almost normally distributed,
which is a good approximation because $\bar{A}$ is a weighted average 
of 50 independent values, 
this expression allows us to determine a 95\%-confidence
interval $\left[ \bar{A}-2 \sqrt{ \mbox{var} (\bar{A})},
        \bar{A}+2 \sqrt{ \mbox{var} (\bar{A})}\right]$
for $\langle \hat{A} \rangle_{\beta}$. 
Expression \ref{avar} also demonstrates that the statistical error
is inversely proportional to the average sign $\mbox{E}_{|w|} (s)$.
\end{itemize}

\section{Results}
\label{sectionresults}
\subsection{Proton and neutron contributions}
\begin{figure}
\centerline{\epsfysize=8cm \epsfbox{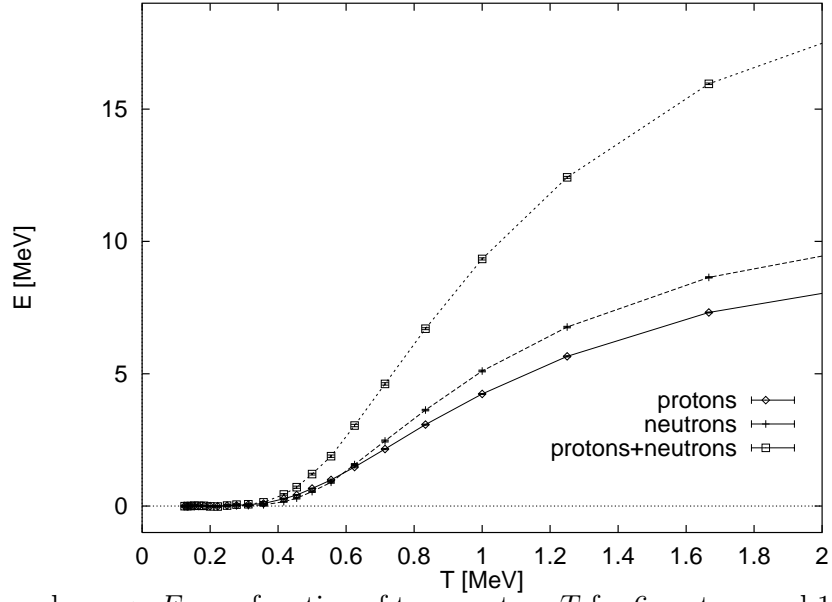}}
\caption{Internal energy $E$ as a function of temperature $T$
         for 6 protons and 10 neutrons in the  
         $ 1f_{\frac{7}{2}} 2p_{\frac{3}{2}} 2p_{\frac{1}{2}} 
         1f_{\frac{ 5}{2}} $ shell, $G=20MeV/56$.
         The energy scale is adapted  such that the total, proton 
         and neutron internal energies all tend to 0 at low temperature.}
\label{figurepairprotonneutrone}
\end{figure}

\begin{figure}
\centerline{\epsfysize=8cm \epsfbox{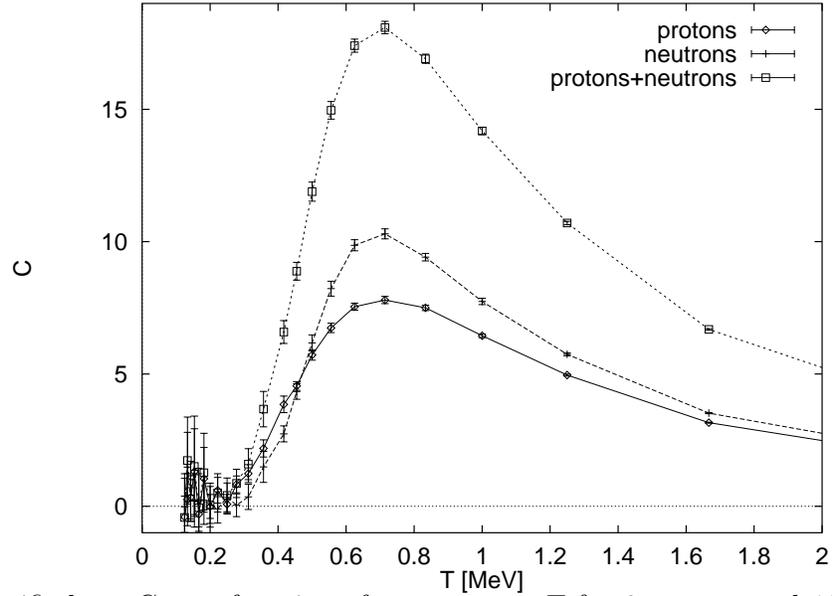}}
\caption{Specific heat $C$ as a function of temperature $T$
         for 6 protons and 10 neutrons in the  
         $ 1f_{\frac{ 7}{2}} 2p_{\frac{ 3}{2}} 2p_{\frac{ 1}{2}} 
         1f_{\frac{ 5}{2}} $ shell, $G=20MeV/56$.}
\label{figurepairprotonneutronc1}
\end{figure}

Some thermodynamical properties of the pairing model for 
$\mbox{}^{56}_{26}\mbox{Fe}_{30}$ were studied using the 
Quantum Monte-Carlo method presented in section \ref{sectionqmc}.
Because the proton and neutron systems are not coupled to one another,
separate results for both particle types are obtained.
The internal energy of the total system
and the contributions of the proton and neutron subsystems 
are shown as a function of temperature in figure 
\ref{figurepairprotonneutrone}.
The same is done for the specific heat in figure 
\ref{figurepairprotonneutronc1}.
The neutrons contribute more to the internal energy than the protons do,
because there are more valence neutrons than valence protons.
This also leads to a slightly stronger peak in the specific-heat curve
for neutrons than for protons.
Qualitatively, there is no big difference between the thermodynamical 
properties of both subsystems.
This is not the case at lower values of the interaction strength $G$.
\subsection{Dependence on the pairing interaction strength G}
We have studied the pairing model for $\mbox{}^{56}_{26}\mbox{Fe}_{30}$
for several values of the pairing interaction strength.
Calculations were performed for $10$ neutrons in a shell with $20$ 
valence states
($1f_{\frac{ 7}{2}} $, $ 2p_{\frac{ 3}{2}} $,$ 2p_{\frac{ 1}{2}} $ 
and $ 1f_{\frac{ 5}{2}} $ orbitals)
and for $6$ protons in the same shell.

\begin{figure}
\centerline{\epsfysize=8cm \epsfbox{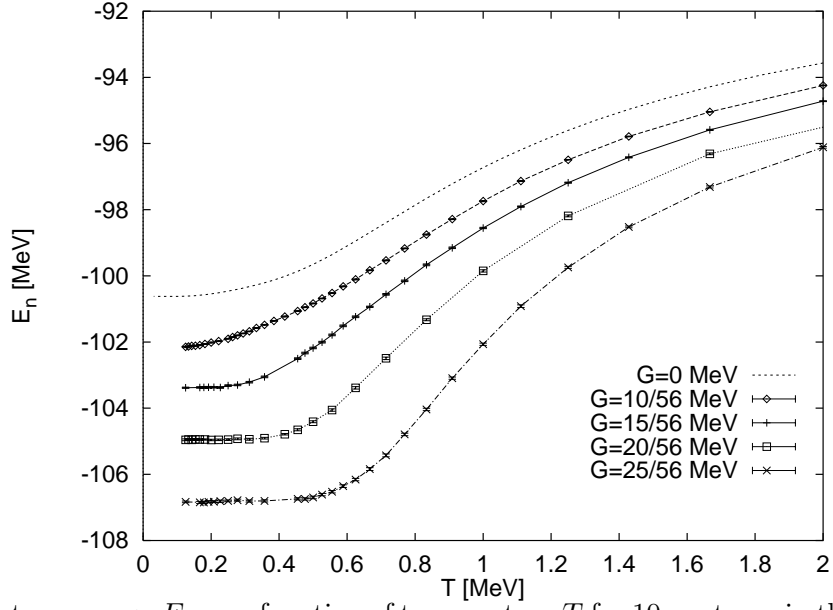}}
\caption{Neutron energy $E_n$ as a function of temperature $T$
         for 10 neutrons in the  
         $ 1f_{\frac{ 7}{2}} 2p_{\frac{ 3}{2}} 2p_{\frac{ 1}{2}} 
         1f_{\frac{ 5}{2}} $ shell,
         for various values of the pairing strength $G$ }
\label{figurepairneutrone}
\end{figure}

\begin{figure}
\centerline{\epsfysize=8cm \epsfbox{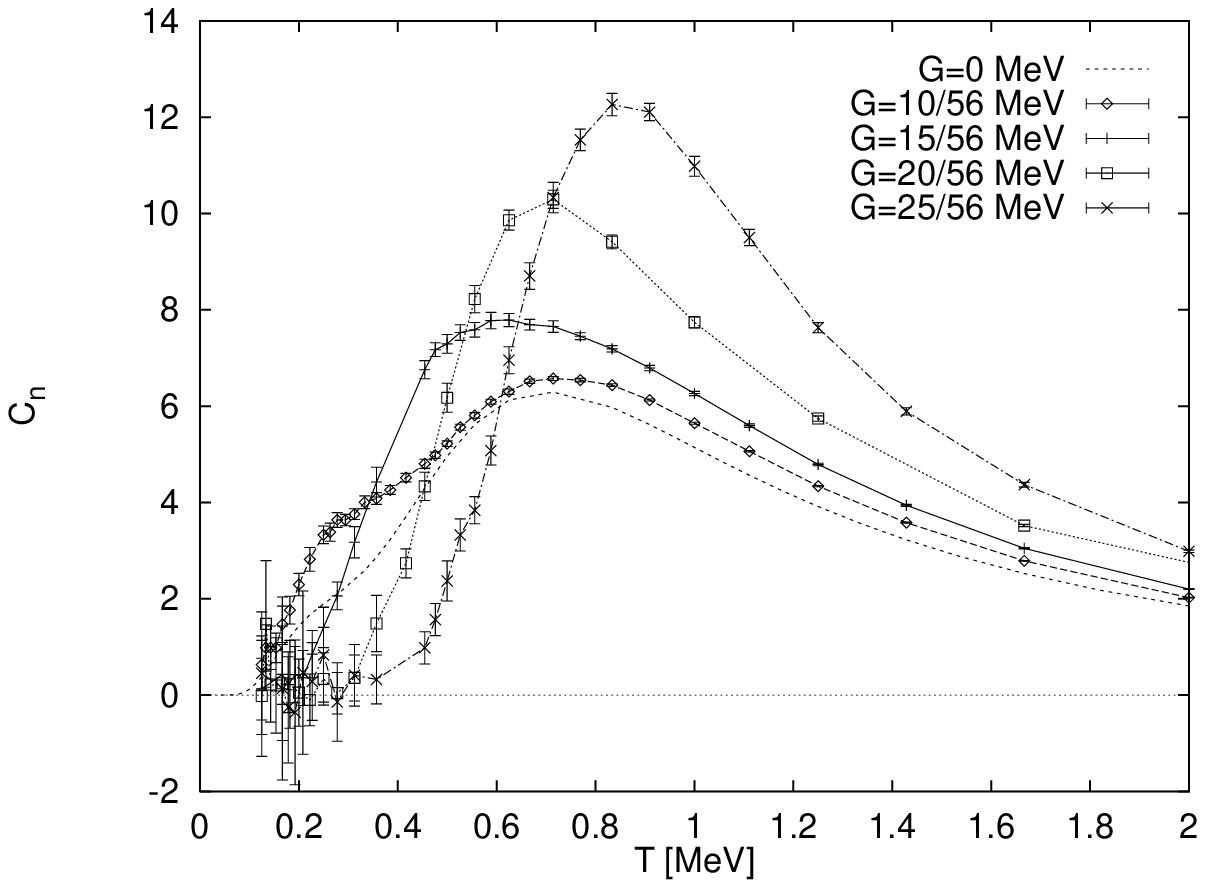}}
\caption{Neutron specific heat $C_n$ as a function of temperature $T$
         for  10 neutrons in the  
         $ 1f_{\frac{ 7}{2}} 2p_{\frac{ 3}{2}} 2p_{\frac{ 1}{2}} 
         1f_{\frac{ 5}{2}} $ shell,
         for various values of the pairing strength $G$ }
\label{figurepairneutronc1}
\end{figure}

The neutron energy as a function of temperature is shown in figure 
\ref{figurepairneutrone}.
The energy scale is chosen such that the inert core has zero energy.
The fact that the energy does not go to much higher values 
as the temperature increases, is due to the limited size of the model space:
not enough high-lying  states are included.
As we shall discuss later on, 
the results for $T \geq 1.5 MeV$ are not physical anymore.
For larger values of $G$, the system is more strongly bound.
Furthermore, when raising the temperature,
the system stays in its ground state longer than for smaller values of $G$.
This indicates that there is an energy gap between the ground state 
and the first excited state proportional to $G$, 
as is expected from BCS theory \cite{ring}.
The neutron specific heat, as a function of temperature, is shown 
in figure \ref{figurepairneutronc1}.
With increasing strength $G$, the peak in the specific heat curve shifts 
to a slightly higher temperature and becomes more pronounced.
In general, peaks in the specific heat can be interpreted 
as signs of a phase transition.
We see here that the pairing correlations, for $G \geq 20MeV/56$, 
seem to induce a phase transition in the system.

\begin{figure}
\centerline{\epsfysize=8cm \epsfbox{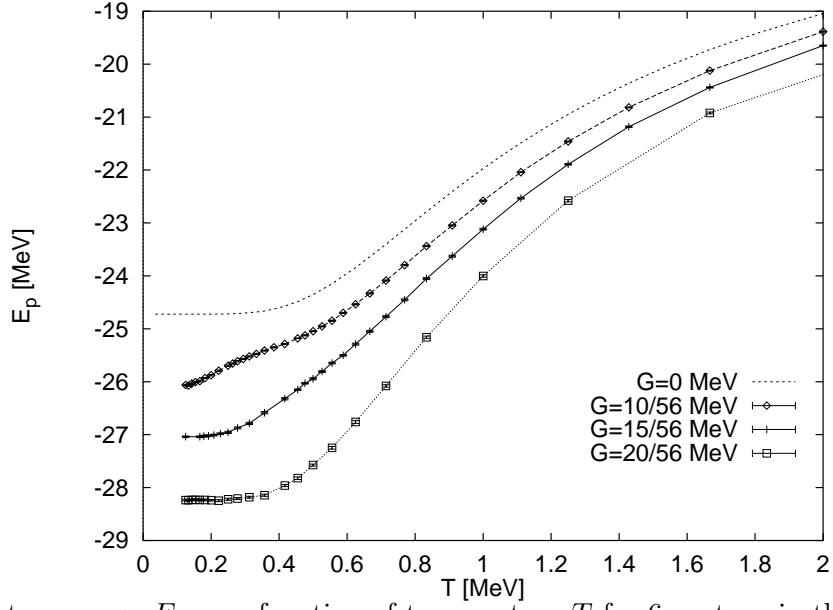}}
\caption{Proton energy $E_p$ as a function of temperature $T$
         for 6 protons in the  
         $ 1f_{\frac{ 7}{2}} 2p_{\frac{ 3}{2}} 2p_{\frac{ 1}{2}} 
         1f_{\frac{ 5}{2}} $ shell,
         for various values of the pairing strength $G$ }
\label{figurepairprotone}
\end{figure}

\begin{figure}
\centerline{\epsfysize=8cm \epsfbox{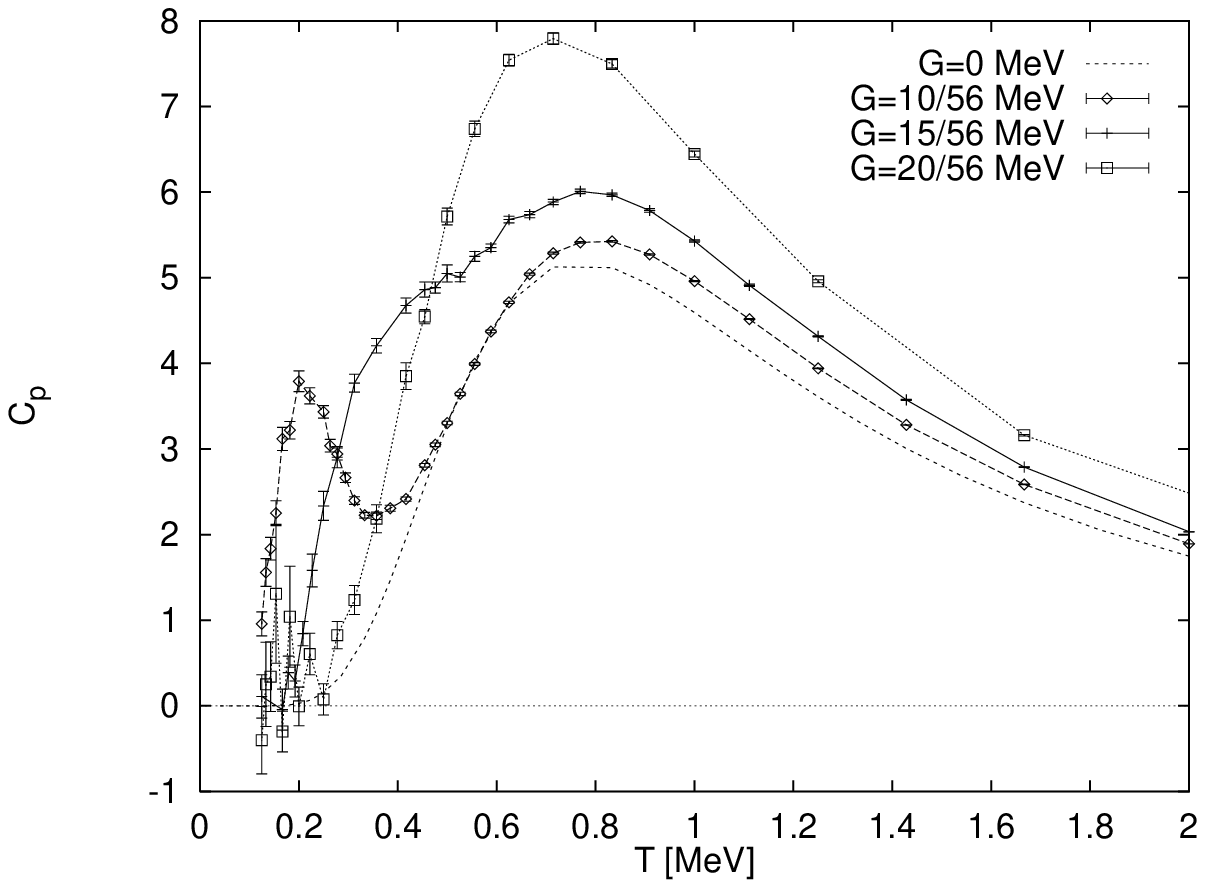}}
\caption{Proton specific heat $C_p$ as a function of temperature $T$
         for 6 protons in the  
         $ 1f_{\frac{ 7}{2}} 2p_{\frac{ 3}{2}} 2p_{\frac{ 1}{2}} 
         1f_{\frac{ 5}{2}} $ shell,
         for various values of the pairing strength $G$ }
\label{figurepairprotonc1}
\end{figure}

\begin{figure}
\centerline{\epsfysize=8cm \epsfbox{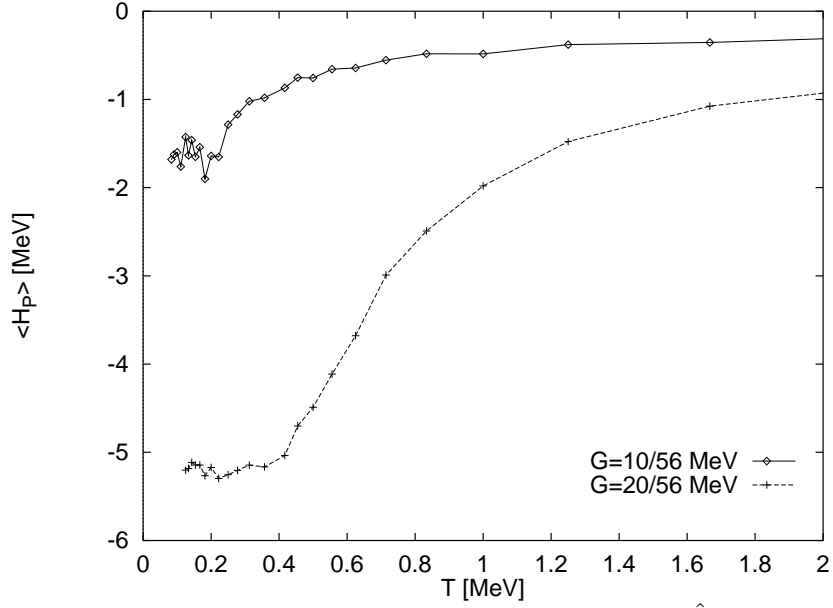}}
\caption{The expectation value of the pairing-interaction operator 
         $\langle \hat{H}_2 \rangle$ as a function of temperature $T$
         for pairing strength $G=10 MeV/56$ and $G=20 MeV/56$ 
         and  6 protons in the  
         $ 1f_{\frac{ 7}{2}} 2p_{\frac{ 3}{2}} 2p_{\frac{ 1}{2}} 
         1f_{\frac{ 5}{2}} $ shell.
         No error limits were determined.}
\label{figurepairprotonvp}
\end{figure}

\begin{figure}
\centerline{\epsfysize=8cm \epsfbox{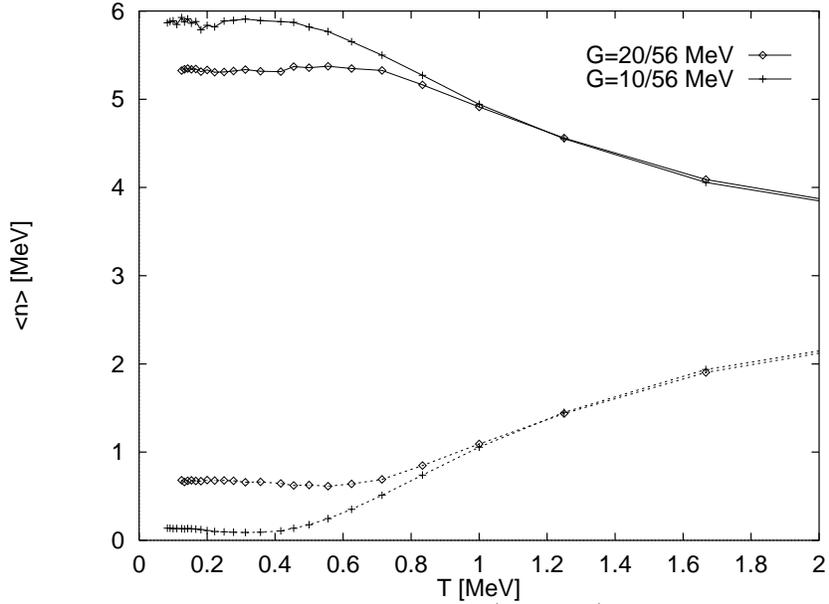}}
\caption{The number particles in the $ 1f_{\frac{ 7}{2}} $ orbital (full line)
         and in the other orbitals (dotted line)
         as a function of the inverse temperature $\beta$,
         for 6 protons in the  
         $ 1f_{\frac{ 7}{2}} 2p_{\frac{ 3}{2}} 2p_{\frac{ 1}{2}} 
         1f_{\frac{ 5}{2}} $ shell.
         No error limits were determined.}
\label{figurepairprotonnf}
\end{figure}

Analogous calculations were done for protons.
The proton energy as a function of temperature is shown 
in figure \ref{figurepairprotone}.
The proton specific heat is shown as a function of temperature 
in figure \ref{figurepairprotonc1}.
The same discussion as for the neutron results, applies here.
There is, however, a striking difference in the specific-heat curve 
for low values of  $G$:
a second peak develops around $T=0.2 MeV$ for $G = 10MeV/56$.
At this value of the pairing strength, 
the broad peak in the specific-heat curve, around $T=0.8 MeV$, 
coincides with the peak in the the specific-heat curve for a pure mean field.
This peak is related to the condensation 
of the valence particles in the lowest energy levels of the valence shell 
(the $ 1f_{\frac{ 7}{2}} $ orbital).
The smaller is entirely due to pair correlations,
that develop among the 6 particles in the $ 1f_{\frac{ 7}{2}} $ orbital.
In figure \ref{figurepairprotonvp}, the expectation value of the 
pairing interaction operator $\hat{H}_P$ is shown as a function of 
the temperature $T$.
While the system with $G= 20MeV/56$ reaches full pairing strength 
at temperatures $T \leq 0.4 MeV$,
the  system with $G= 10MeV/56$ comes to this regime only at values 
of $T \leq 0.2 MeV$.
In figure \ref{figurepairprotonnf}, 
the number particles in the $ 1f_{\frac{ 7}{2}} $ orbital
and the number of particles in the other orbitals are shown.
For the system with $G= 10MeV/56$, it is observed that 
approximately all 6 particles occupy states 
in the $ 1f_{\frac{ 7}{2}} $ orbital for values of $T \leq 0.45 MeV$.
The fact that the pairing correlations reach their maximum 
for this system only at values of $T \leq 0.2 MeV$,
means that the system passes through two phases as it is cooled:
first, the 6 valence protons condense into the $ 1f_{\frac{ 7}{2}} $ orbital.
At $T \simeq 0.45 MeV$, this stage is completed.
If the temperature is lowered further, 
pair correlations among these particles can develop.
At values of $T \leq 0.2 MeV$ the system is almost completely 
cooled to its ground state.
For the system with $G= 20MeV/56$, the occupation 
of the $ 1f_{\frac{ 7}{2}} $ orbital reaches a maximum of about 5.3. 
The particles always remain spread over all the valence orbitals,
because now the pairing interaction is strong enough to scatter them 
out of the $ 1f_{\frac{ 7}{2}} $ orbital, even in the ground state.

\subsection{Dependence on the size of the model space}
For the description of the high-temperature properties of the system,
the model space given by the $fp$ shell is too small.
At temperatures of a few MeV, 
valence particles can be excited to higher-lying single-particle states,
or core particles can be excited into the valence orbitals or higher energy 
states.
In order to know up to what temperatures the results 
that we obtained in the $fp$ shell are valid,
we performed a number of calculations in larger model spaces. 
First, the $3s_{\frac{ 1}{2}}$, $2d_{\frac{ 5}{2}}$, 
$2d_{\frac{ 3}{2}}$ and $1g_{\frac{ 9}{2}}$ orbitals
are added to the single-particle space.
This leads to a many-body problem of 6 and 10 particles 
in 42 single-particle states.
In a second extended model the core states are considered 
as valence states too.  
Therefore the $1s_{\frac{ 1}{2}}$, $1p_{\frac{ 3}{2}}$, 
$1p_{\frac{ 1}{2}}$, $1d_{\frac{ 5}{2}}$, and
$1d_{\frac{ 3}{2}}$, $2s_{\frac{ 1}{2}}$ are added.
Furthermore, also the $3p_{\frac{ 3}{2}}$, $3p_{\frac{ 1}{2}}$, 
$4s_{\frac{ 1}{2}}$ and $1g_{\frac{ 7}{2}}$   
orbitals are taken into account.
This leads to a many-body problem of 26 and 30 particles 
in 78 single-particle states.

Because multiple shells are used, 
the model space of the extended systems contains
spurious excitations related to center-of-mass motion.
Therefore, care has to be taken when relating high temperature results
to internal excitations of the system.
For the largest model space (without core),
these center-of-mass motions can be interpreted 
as thermal excitations of the collective degrees of freedom.
This picture would be physically meaningful 
in the absence of a mean-field potential.
The fact that the mean-field potential is localized in space, 
breaks the translational invariance of the model.
Therefore, one cannot separate the center-of-mass motion 
from the intrinsic excitations in a clean way \cite{lawson}.
A consistent treatment of spurious states is a topic for further research.

\begin{figure}
\centerline{\epsfysize=8cm \epsfbox{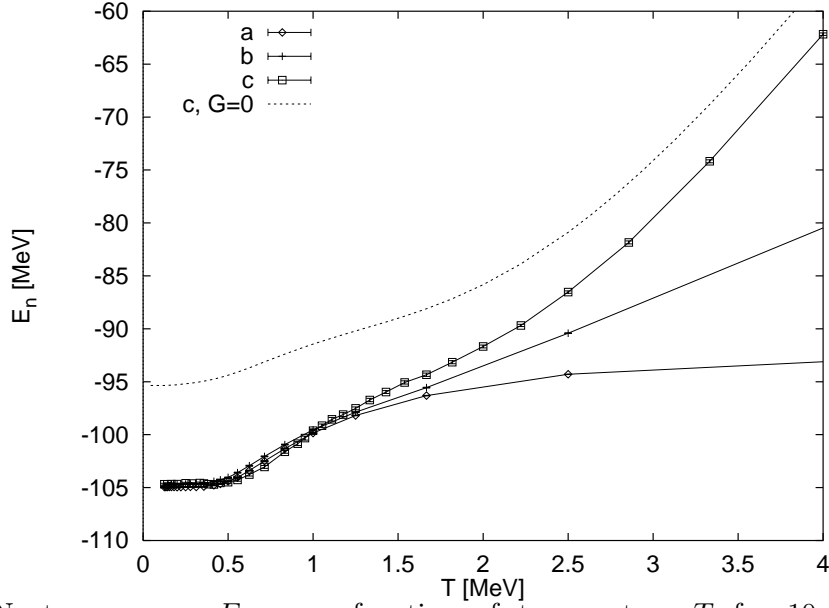}}
\caption{Neutron energy $E_n$ as a function of temperature $T$
         for 10 neutrons in the  
         $ 1f_{\frac{ 7}{2}} 2p_{\frac{ 3}{2}} 2p_{\frac{ 1}{2}} 
         1f_{\frac{ 5}{2}} $ shell (a),
         for 10 neutrons in the first extended model space (b)
         and for 30 neutrons in the second extended model space 
         (c).
	 The dashed line gives the result 
         for the second extended model space without pairing ($G=0$).
	 The results for the largest modelspace are shifted such 
         that curve (b) and (c) coincide at low temperature}
\label{figurepairneutronxe}
\end{figure}

\begin{figure}
\centerline{\epsfysize=8cm \epsfbox{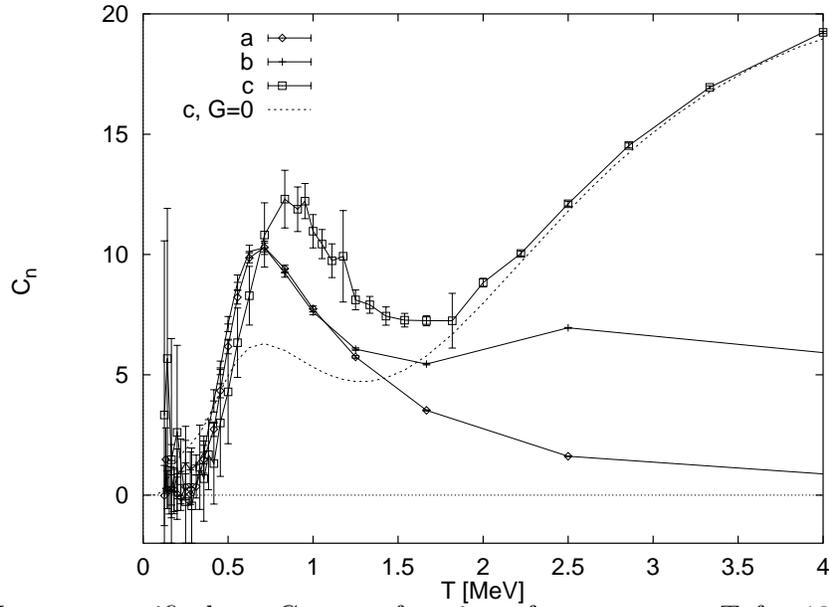}}
\caption{Neutron specific heat $C_n$ as a function of temperature $T$
         for  10 neutrons in the  
         $ 1f_{\frac{ 7}{2}} 2p_{\frac{ 3}{2}} 2p_{\frac{ 1}{2}} 
         1f_{\frac{ 5}{2}} $ shell (a),
         for  10 neutrons in the first extended model space (b)
         and for  30 neutrons in the second extended model space (c).
	 The dashed line gives the result 
         for the second extended model without pairing ($G=0$).}
\label{figurepairneutronxc1}
\end{figure}

\begin{figure}
\centerline{\epsfysize=8cm \epsfbox{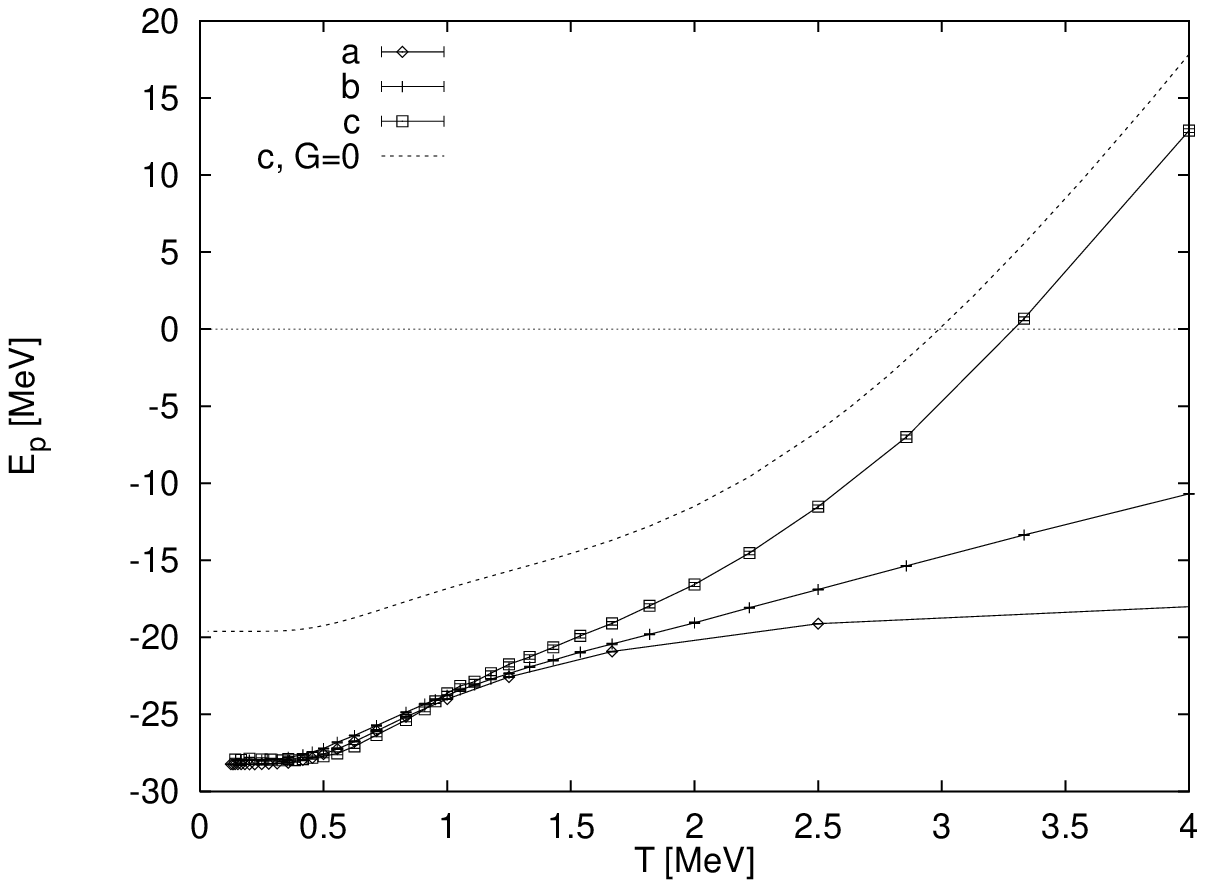}}
\caption{Proton energy $E_p$ as a function of temperature $T$
         for  6 protons in the  
         $ 1f_{\frac{ 7}{2}} 2p_{\frac{ 3}{2}} 2p_{\frac{ 1}{2}} 
         1f_{\frac{ 5}{2}} $ shell (a),
         for  6 protons in the first extended model space (b)
         and for 26 protons in the second extended model space (c).
	 The dashed line gives the result 
         for the second extended model without pairing ($G=0$).
	 The results for the largest modelspace are shifted such 
         that curve (b) and (c) coincide at low temperature}
\label{figurepairprotonxe}
\end{figure}

\begin{figure}
\centerline{\epsfysize=8cm \epsfbox{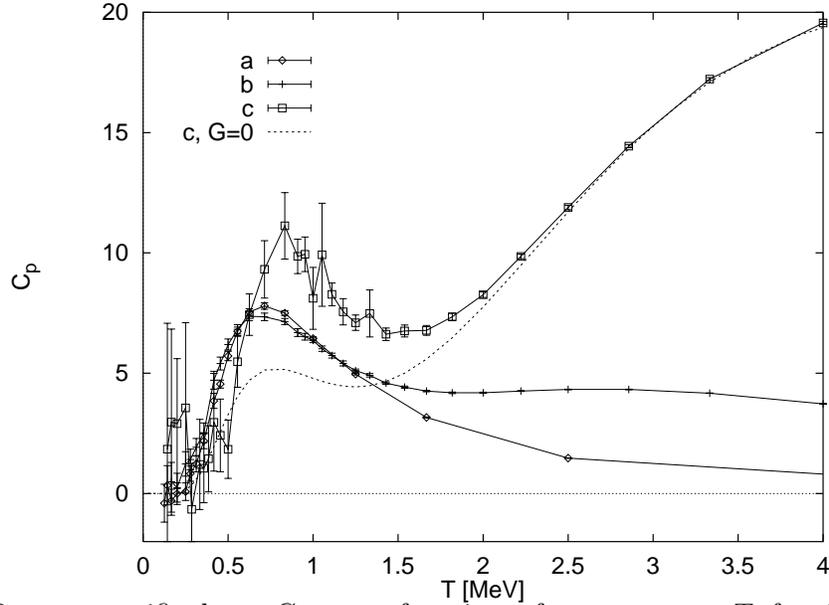}}
\caption{Proton specific heat $C_p$ as a function of temperature $T$
         for 6 protons in the  
         $ 1f_{\frac{ 7}{2}} 2p_{\frac{ 3}{2}} 2p_{\frac{ 1}{2}} 
         1f_{\frac{ 5}{2}} $ shell (a),
         for 6 protons in the first extended model space (b)
         and for 26 protons in the second extended model space (c).
	 The dashed line gives the result 
         for the second extended model without pairing ($G=0$).}
\label{figurepairprotonxc1}
\end{figure}

The results for the energy and the specific heat 
obtained using these model spaces
are shown in figure \ref{figurepairneutronxe} to \ref{figurepairprotonxc1}.
If the value for the pairing interaction strength $G$ is not changed,
then a system with a larger model space will have a lower ground-state energy  
because the larger model space allows stronger pair correlations.
In order to obtain a comparable pairing energy, a reduced pairing 
interaction strength of $G= 16 MeV/56$ is used for the extended model spaces.
For the no-core system, the energy is shifted such that the ground-state energy
coincides with the ground-state energy of the $fp$ shell system. 

In the largest model space, at high temperatures ($T \geq 2 MeV$),  
the specific-heat curve coincides with the specific-heat curve for $G=0$.
In this temperature region,
the proton and neutron energy are some $5 MeV$ 
lower than in the $G=0$ case.
Apart from this shift, the energy curves 
are similar to the $G=0$ case.
This indicates that, at high temperatures, 
the pairing Hamiltonian enhances the binding energy   
but has no effect on the internal structure.

At lower temperatures, the specific heat curve deviates 
from the curve for $G=0$, because pairing correlations develop.
By comparing the results for the $fp$ shell and the first extended model space,
we see that even the $fp$ shell is too small 
to describe the system at temperatures $T \geq 1.3 MeV$.
In order to compare with the results for the second extended model space 
around temperatures of $1 MeV$, the pairing interaction strength $G$ ought 
to be reduced somewhat more for the latter model space.
The vanishing of pair correlations with increasing temperature, starting 
from $T \simeq 1 MeV$,
was also observed in shell-model quantum Monte-Carlo calculations for  
$\mbox{}^{54}_{26}\mbox{Fe}_{28}$
based on more realistic interactions \cite{deanfe,pairsmmc1}.
The interesting topic of proton-neutron pairing in $N=Z$ nuclei 
\cite{pairsmmc2}, could of course not be adressed 
in our too schematic model used at present.

The schematic form of the interaction also urges for a caveat:
in the larger model spaces, the pairing interaction scatters
particles into Woods-Saxon orbitals that extend into regions
where the mean field vanishes (the orbitals beyond the core and the fp shell).
Because the pairing interaction introduces correlations in the angular
coordinates but not in the radial coordinates, one might expect
the ground state wavefunction to extend too far into space.
A more realistic interaction would localize the wavefunction more
closely to the core.
Our calculations show that these extended orbitals 
contribute only about 1 \% of the ground state density.
Therefore we think that this has only a minor effect
on the energy-like observables we are interested in here.

\subsection{Dependence on the number of particles}
\begin{figure}
\centerline{\epsfysize=8cm \epsfbox{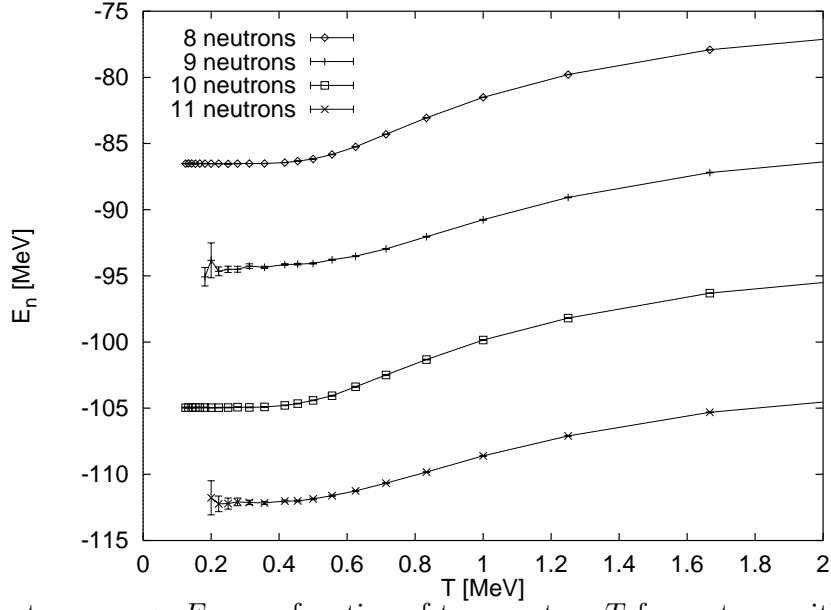}}
\caption{Neutron energy $E_n$ as a function of temperature $T$
         for systems with 8, 9, 10 and 11 neutrons in the  
         $ 1f_{\frac{ 7}{2}} 2p_{\frac{ 3}{2}} 2p_{\frac{ 1}{2}} 
         1f_{\frac{ 5}{2}} $ shell, ($G=20MeV/56$). }
\label{figurepairnu}
\end{figure}

\begin{figure}
\centerline{\epsfysize=8cm \epsfbox{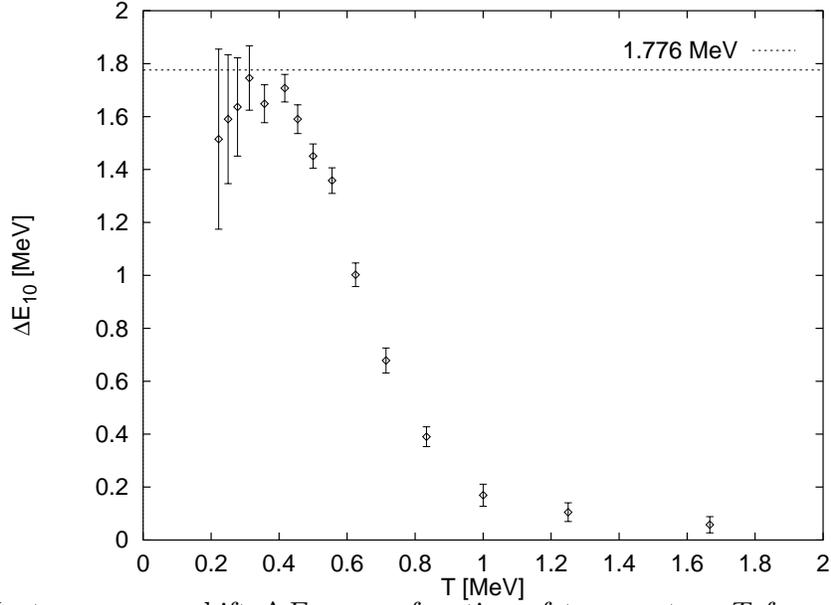}}
\caption{Neutron energy shift ${\Delta E}_{10}$ 
         as a function of temperature $T$
         for a systems with 10 neutrons in the  	 
         $ 1f_{\frac{ 7}{2}} 2p_{\frac{ 3}{2}} 2p_{\frac{ 1}{2}} 
         1f_{\frac{ 5}{2}} $ shell, ($G=20MeV/56$). 
         The dashed line indicates the experimental value 
         of the ground-state energy shift 
         for $\mbox{}^{56}_{26}\mbox{Fe}_{30}$.}
\label{figurepairndu}
\end{figure}

We studied systems with various numbers of neutrons in the $fp$ shell:
$\mbox{}^{54}_{26}\mbox{Fe}_{28}$, $\mbox{}^{55}_{26}\mbox{Fe}_{29}$, 
$\mbox{}^{56}_{26}\mbox{Fe}_{30}$
and $\mbox{}^{57}_{26}\mbox{Fe}_{31}$ are modelled by considering 
$8$, $9$, $10$ and $11$ neutrons
in the $fp$ valence shell, respectively.
For the systems with $9$ and $11$ neutrons,
the sign problem limits accurate calculations to values of 
$T \geq 0.25 {MeV}$.
Fortunately, this temperature is low enough to get a good approximation 
of the ground state.
The neutron energy $E_n$ for the various systems is shown 
as a function of temperature in figure \ref{figurepairnu}.
The proton internal energy is not shown because it is equal 
for all four systems 
and it is already given in  figure  \ref{figurepairprotone}.
While at high temperature the energy curves are equidistantly spaced,
with an interval of about $9 MeV$,
there is a relative shift to lower energies 
for the systems with $8$ and $10$ neutrons at low energy.
This is because the pairing correlations are stronger for the even systems 
than for the odd systems at temperatures below $1 MeV$.
The shift in the energy for the system with 10 neutrons can be 
quantified as 
\begin{equation}
 {\Delta E}_{10} = \frac{ E_{n 9} +E_{n 11}}{2} -E_{n 10},
\label{intushift}
\end{equation}
with $E_{n 9}$, $E_{n 10}$, $E_{n 11}$ the neutron energies for the systems
with 9, 10 and 11 valence neutrons respectively.
The quantity ${\Delta E}_{10}$ is shown as a function of temperature in 
figure \ref{figurepairndu}.
The ground-state energy shift was calculated analogously to expression 
\ref{intushift},
with the energies replaced by the mass excesses given in reference \cite{tuli}.
A value of $1.776 MeV$ is obtained.
Our quantum Monte-Carlo results approach this value remarkably well 
at temperatures below $0.5 MeV$.

\subsection{Level densities}
\begin{figure}
\centerline{\epsfysize=8cm \epsfbox{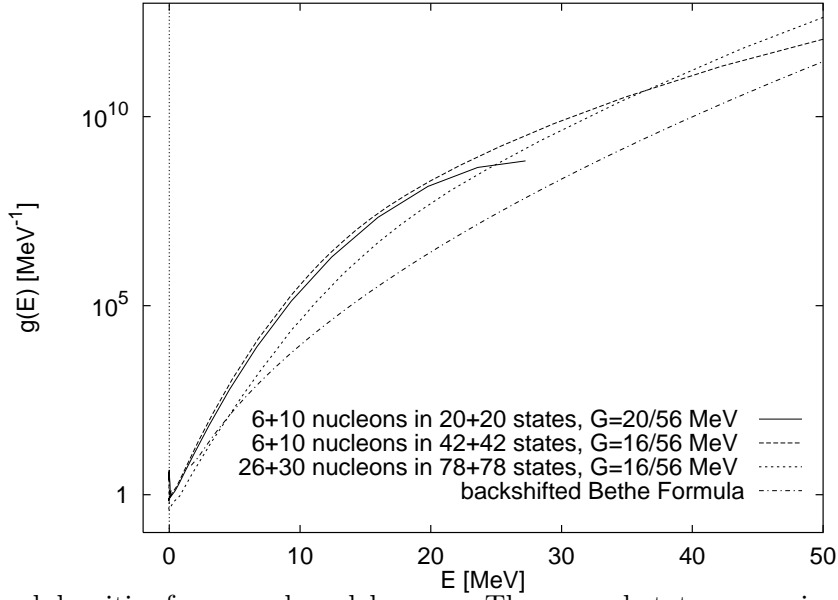}}
\caption{Level densities for several model spaces.
         The ground-state energy in each model space was shifted to 0 MeV.
         The backshifted Bethe formula was fitted to experimental data 
         for $\mbox{}^{56}Fe$.} (for temperatures up to $10^{10}$ K).
\label{figurelevel1}
\end{figure}

Because the internal energy is related to the derivative 
of the logarithm of the partition function $Z_{\beta}$ 
and because $Z_{\beta}$ is the Laplace transform 
of the level density $g(E)$ of excited states,
the results presented above also give information about the level density.
The partition function can be obtained by numerical integration 
of the internal energy. 
Then one should apply an inverse laplace transform on $Z_{\beta}$.
Because of the statistical errors on the internal energies 
and hence on $Z_{\beta}$, this is however an ill-posed problem.
A good approximation (at high enough energies) is given by the saddle-point
approximation:
\begin{equation}
g(E)=\frac{e^{\beta E} Z_{\beta}} {\sqrt{2 \pi (\beta^{-2}C)}}.
\end{equation}
Our quantum Monte-Carlo method gives accurate results for $E$ and 
$\beta^{-2}C$. 
So the level density in the  saddle-point approximation is easily obtained.
Figure \ref{figurelevel1} shows the level densities derived from the 
results shown in figures 
 \ref{figurepairneutronxe}, \ref{figurepairneutronxc1}, 
 \ref{figurepairprotonxe} and \ref{figurepairprotonxc1}.
The results for the smallest model space only go up to excitation energies 
of 25 MeV. 
They are in good agreement with the results 
for the first extended model space, up to energies of 20 MeV. 
However, the results for the largest model space deviate from these, 
even at energies below 20 MeV. 
This indicates that it is important to consider also core excitations
when calculating level densities.
Figure  \ref{figurelevel1} also shows the level-density curve 
from a backshifted Bethe formula cited in reference \cite{nakada},
with parameters $a=5.80 {MeV}^{-1}$ and $\Delta=1.38 MeV$.
This parametrization  was fitted to experimental data \cite{woosley}
in order to reflect finite-temperature properties at temperatures
between $10^7$ an $10^{10} K$.
Clearly, the level densities are shited too much to lower energies
by the residual interaction we considered.
Shell-model Monte-Carlo calculations with a more realistic interaction,
but in a limited model space 
(the $1f_{\frac{ 7}{2}} 2p_{\frac{ 3}{2}} 2p_{\frac{ 1}{2}} 1f_{\frac{ 5}{2}} 
 1g_{\frac{ 9}{2}}$ shell),
result in a better agreement with the backshifted Bethe formula
at energies between 5 and 20 MeV \cite{nakada}.

\begin{figure}
\centerline{\epsfysize=8cm \epsfbox{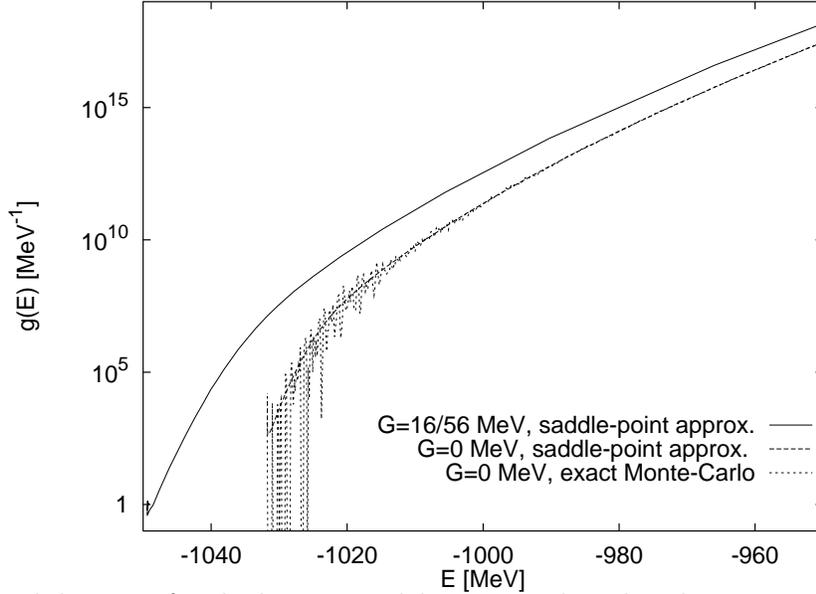}}
\caption{Level densities for the largest model space, 
         with and without pairing interaction.
         The energy scale reflects the sum of the single-particle energies
         as listed in table \ref{table1} and the additional binding energy
         stemming from the pairing interaction.}
\label{figurelevel2}
\end{figure}

Figure \ref{figurelevel2} compares the level density for the largest model 
space with the level density for the mean-field Hamiltonian.
For the mean-field case, the level density can be calculated exactly 
using a Monte-Carlo method devised by Cerf \cite{cerfl}. 
We performed such a calculation for the largest model space in order to 
compare it with the saddle-point approximation.
For this largest model space, the saddle-point results follow
closely the exact level density at high energies. 
At lower energies, the exact level density shows a structure with a lot of
peaks, because of the discrete structure of the spectrum.
In the saddle-point approximations these peaks are absent.
If these peaks are smeared out with a width of 0.5 MeV, 
the curve coincides with the saddle-point approximation
even at excitation energies as low as 1 MeV.
Thus the saddle-point approximation gives a good 'smoothed' estimate.
In the smaller model spaces the agreement is less good.
Figure \ref{figurelevel2} shows that the residual interaction shifts
level densities to lower energies. 
This shift is largest at low energies, because there the pairing correlations
are strongest.

\section{Conclusion}
We conclude by stating that our Quantum Monte-Carlo method offers 
a powerful tool for the study of the nuclear pairing model.
We have put emphasis on the thermodynamical properties.
Occupation numbers and the pairing gap can be calculated too using this
method.
Main advantages over other methods are that many-body correlations 
are taken into account exactly,
particle numbers are constant and finite temperature results can be obtained.
The major disadvantage of the method is that spectroscopic information can 
only be obtained indirectly.
Finally, we remark that our calculations 
indicate that pairing correlations are important
only at low temperature  ( below $1 MeV$) and at low excitation energies,
though they do enhance the binding energy.
A signal of a phase transition related to pairing is found at
temperatures around $0.7 MeV$.
The pairing interaction also shifts the level density towards lower energies.
Furthermore our results show that it is necessary to work in very
large model spaces, that also include core excitations,
if one wants to calculate accurate level densities.
The work presented here constituted a part of the Ph.D. thesis 
of S. Rombouts \cite{thesis}. 

\section*{Acknowledgements}
Discussions with K. Langanke, C.W. Johnson, W.E. Ormand and S. Goriely
are gratefully acknowledged.
The authors are grateful to the F.W.O. (Fund for Scientific Research) 
- Flanders and to the Research Board of the University of Gent for
financial support. 
One of the authors (K.H.) is grateful to CERN for financial support.

\newpage
\begin{table}
\begin{tabular}{|c||r@{\hspace{1.5cm}}|r@{\hspace{1.5cm}}|}
\hline
 orbital          &   \multicolumn{2}{c|} {single-particle energies (MeV)} \\
\cline{2-3}
     \hspace{3cm} &   \hspace{1.5cm} protons &   \hspace{1.5cm}   neutrons \\
\hline
$ 1s_{\frac{ 1}{2}} $ &  -34.7106  &    -42.0333    \\  
$ 1p_{\frac{ 3}{2}} $ &  -25.3351  &    -32.2120    \\
$ 1p_{\frac{ 1}{2}} $ &  -24.0715  &    -31.1979    \\
$ 1d_{\frac{ 5}{2}} $ &  -15.0034  &    -21.5607    \\
$ 1d_{\frac{ 3}{2}} $ &  -12.7911  &    -19.6359    \\
$ 2s_{\frac{ 1}{2}} $ &  -12.3511  &    -19.1840    \\
\hline
$ 1f_{\frac{ 7}{2}} $ &   -4.1205  &    -10.4576    \\
$ 2p_{\frac{ 3}{2}} $ &   -2.0360  &     -8.4804    \\  
$ 2p_{\frac{ 1}{2}} $ &   -1.2334  &     -7.6512    \\  
$ 1f_{\frac{ 5}{2}} $ &   -1.2159  &     -7.7025    \\  
\hline
$ 3s_{\frac{ 1}{2}} $ &    4.7316  &     -0.3861    \\  
$ 2d_{\frac{ 5}{2}} $ &    5.6562  &      0.2225    \\  
$ 2d_{\frac{ 3}{2}} $ &    6.1324  &      0.9907    \\  
$ 1g_{\frac{ 9}{2}} $ &    6.6572  &      0.5631    \\  
$ 3p_{\frac{ 3}{2}} $ &    6.6663  &      2.5931    \\ 
$ 3p_{\frac{ 1}{2}} $ &    6.7469  &      2.6915    \\ 
$ 4s_{\frac{ 1}{2}} $ &    8.9016  &      4.4706    \\ 
$ 1g_{\frac{ 7}{2}} $ &    9.1386  &      3.5488    \\ 
\hline						   
\end{tabular}
\caption{Single-particle eigenstates of the Woods-Saxon potential
         with the parameters as described in section \ref{sectionmodel}}
\label{table1}
\end{table}

\begin{figure}
\centerline{\epsfysize=8cm \epsfbox{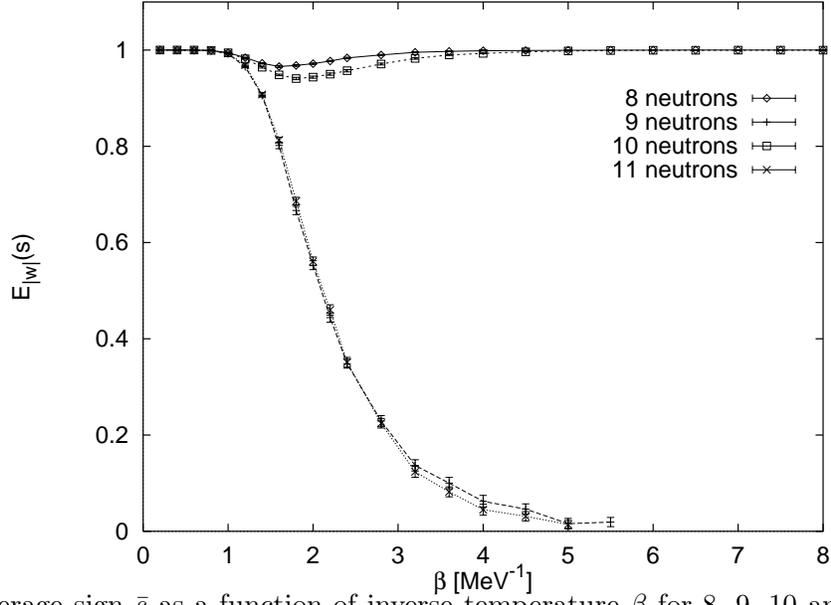}}
\caption{Average sign $\bar{s}$ as a function of inverse temperature $\beta$
         for 8, 9, 10 and 11 neutrons in the  
         $ 1f_{\frac{ 7}{2}} 2p_{\frac{ 3}{2}} 2p_{\frac{ 1}{2}} 
         1f_{\frac{ 5}{2}} $ shell, $G=20MeV/56$.}
\label{figurepairnsigns}
\end{figure}
\newpage
\end{document}